\documentclass[fleqn,usenatbib]{mnras}

\usepackage{newtxtext,newtxmath}
\usepackage{xcolor}
\usepackage[T1]{fontenc}
\usepackage{subfig}
\usepackage{dcolumn}

\newcolumntype{L}{D{.}{.}{2,2}}

\DeclareRobustCommand{\VAN}[3]{#2}
\let\VANthebibliography\thebibliography
\def\thebibliography{\DeclareRobustCommand{\VAN}[3]{##3}\VANthebibliography}

\usepackage{graphicx}	% Including figure files
\usepackage{amsmath}	% Advanced maths commands
\usepackage{gensymb}
\defcitealias{Posselt2023}{P23}
\defcitealias{Spiewak2022}{S22}
\defcitealias{Szary14}{S14}

\title[Emission properties of slow pulsars and MSPs]{The Thousand-Pulsar-Array programme on MeerKAT XV: A comparison of the radio emission properties of slow and millisecond pulsars}

\author[A. Karastergiou et al.]{
A. Karastergiou\,$^{1}$,\thanks{E-mail: aris.karastergiou@physics.ox.ac.uk}
S. Johnston\,$^{2}$,
B. Posselt\,$^{1}$,
L. S. Oswald\,$^{1,3}$,
M.~Kramer\,$^{4,5}$,
P. Weltevrede\,$^{5}$
\\
$^{1}$ Department of Astrophysics, University of Oxford, Denys Wilkinson Building, Keble Road, Oxford OX1 3RH, UK\\
$^2$Australia Telescope National Facility, CSIRO, Space and Astronomy, PO Box 76, Epping, NSW 1710, Australia.\\
$^{3}$ Magdalen College, University of Oxford, Oxford OX1 4AU, UK\\
{$^{4}$ Max-Planck-Institut f\"{u}r Radioastronomie, Auf dem H\"{u}gel 69, D-53121 Bonn, Germany}\\
{$^{5}$ Jodrell Bank Centre for Astrophysics, Department of Physics and Astronomy, University of Manchester, Manchester M13 9PL, UK}\\
}

\date{Accepted XXX. Received YYY; in original form ZZZ}

\pubyear{2015}
\def\ppdot{\,$P-\dot{P}$\,}
\def\bsurf{\,$B_{\rm{surf}}$\,}
\def\tauc{\,$\tau_{\rm c}$\,}
\def\Smean{\,$\bar{S}_\nu$\,}
\def\Smeanm{\,\bar{S}_\nu\,}
\begin{document}
\label{firstpage}
\pagerange{\pageref{firstpage}--\pageref{lastpage}}
\maketitle

\begin{abstract}
We use data from the MeerTime project on the MeerKAT telescope to ask whether the radio emission properties of millisecond pulsars (MSPs) and slowly rotating, younger pulsars (SPs) are similar or different. We show that the flux density spectra of both populations are similarly steep, and the widths of MSP pulsar profiles obey the same dependence on the rotational period as slow pulsars. We also show that the polarization of MSPs has similar properties to slow pulsars. The commonly used pseudo-luminosity of pulsars, defined as the product of the flux density and the distance squared, is not appropriate for drawing conclusions about the relative intrinsic radio luminosity of SPs and MSPs. We show that it is possible to scale the pseudo-luminosity to account for the pulse duty cycle and the solid angle of the radio beam, in such a way that MSPs and SPs do not show clear differences in intrinsic luminosity. The data, therefore, support common emission physics between the two populations in spite of orders of magnitude difference in their period derivatives and inferred, surface, dipole magnetic field strengths.
\end{abstract}


%%%%%%%%%%%%%%%%%%%%%%%%%%%%%%%%%%%%%%%%%%%%%%%%%%

%%%%%%%%%%%%%%%%% BODY OF PAPER %%%%%%%%%%%%%%%%%%

\section{Introduction and motivation}
A key tool in the interpretation of the population of radio emitting neutron stars is the \ppdot{} diagram, where the rotational period $P$ is plotted against its first time derivative $\dot{P}$. The rationale has always been that, assuming pulsars spin down due to the fact that they are rotating misaligned magnetic dipoles, one can easily draw on these diagrams lines of constant surface magnetic field \bsurf$\propto\sqrt{P\dot{P}}$ and lines of constant characteristic age  \tauc$= P/2\dot{P}$. Furthermore, one can draw lines of constant $\dot{E}\propto\dot{P}/P^3$, denoting the rate of loss of rotational kinetic energy, which sets the power available for any physical process that ultimately generates the radio emission. There is then an expectation that these inferred physical properties must, in some way, govern pulsar phenomenology. Stated differently, there is an expectation that pulsars in any given region of the \ppdot{} diagram will share physical characteristics and therefore produce similar observable phenomena. 

This expectation is met only in part. While there are confirmed correlations between emission properties and $\dot{E}$, strong correlations with \bsurf and \tauc are not seen. As an example, Soft-Gamma-Repeaters and Anomalous-Xray-Pulsars, formerly designated as such due to the specific phenomena their names imply, occupy at least 3 orders of magnitude in \bsurf space. Even more emphatically, the entire population of neutron stars that can be placed on the \ppdot diagram at present, spans a range of over 7 orders of magnitude in \bsurf, despite exhibiting phenomena which are broadly similar. On the other hand, $\dot{E}$ has been shown to be well correlated with 2 observable quantities, namely the radio luminosity and the degree of linear polarization, as reported in \cite{avh98}, \cite{wj08}, \cite{jk18} and \citet{Posselt2023}. $\dot{E}$ has also been shown to correlate with emission of high energy X-rays and $\gamma$-rays \citep{Smith23}, the latter correlation connecting the population of high-$\dot{E}$ recycled millisecond pulsars (MSPs) with the population of high-$\dot{E}$, young, non-recycled pulsars, despite their inferred \bsurf values differing by 3.5 orders of magnitude. 

This brings us to the core question we address in this paper. Clearly MSPs occupy a defined space in the \ppdot{} diagram. Observationally, are the properties of radio emission from MSPs different to those of slow pulsars? We adopt the term SPs for the slow, younger pulsar population that have not been recycled. This question was notably posed in \cite{kramer98} and \cite{xilouris98}. In summary, those authors concluded that the main phenomenological differences can be found in the luminosity, pulse width (with multiple pulse components appearing in MSPs), polarization, with MSPs apparently showing flat position angle (PA) profiles, and frequency evolution of the polarized emission. The flux density spectral index and profile complexity do not differ to first order between the populations. The differences in the radio luminosity $L_R$, and in the radio emission efficiency defined as $\xi = L_R/\dot{E}$, are also discussed in \cite{Szary14}.

The recent publication of catalogues from large surveys of SPs and MSPs with the MeerKAT telescope provides the data to conduct a population-wide comparison. In the following, we concentrate on the measured flux density spectral index distributions, the pulse profile widths, linear polarization, and the radio luminosity. In Section \ref{methods} we discuss differences in the methodologies in obtaining the catalogue parameters, and then proceed to compare the MeerKAT survey results in Section \ref{results}. In Section \ref{scaling}, we describe possible ways to estimate the true radio luminosity by carefully considering how beaming changes systematically across the pulsar population. The distance to each pulsar plays a crucial role in any interpretation of radio luminosity. We discuss how distance affects our understanding of SP and MSP luminosities in Section \ref{distance}. Finally, we outline the implications and limitations of our results in Section \ref{conclusions}.

\section{A comparison of the methodology used for the MSP and SP catalogue}\label{methods}
For our comparison, we consider two large surveys of SPs and MSPs, carried out with the 64-dish SARAO MeerKAT telescope with the same instrument setup and similar data processing. A census of 189 MSPs was presented by \citet{Spiewak2022}, hereafter \citetalias{Spiewak2022}, while a census of 1170 SPs was presented by \citet{Posselt2023}, hereafter \citetalias{Posselt2023}.The sky distribution of MSPs, defined as pulsars with period $P<50$~ms and period derivative $\dot{P}<2\times 10^{-17}$~s~s$^{-1}$, is given in Fig. 2 of \citetalias{Spiewak2022}, and the sky distribution of SPs is given in Fig. 4 of \citetalias{Posselt2023}.
The data for both works were obtained in the framework of the MeerTime project \citep{Bailes2020}, a MeerKAT Large Survey Project.
% has four major themes. These include the Thousand Pulsar Array (Johnston et al.2020), the Relativistic Binary programme (Kramer et al. 2021), the Globular Cluster pulsar timing programme (Ridolfi et al. 2021), and the MeerTime Pulsar Timing Array (MPTA),
All observations were centred at a frequency of $\nu=1284$\,MHz in the L-band receiver, and the presented data are restricted to a bandwidth of 775\,MHz.
The scintillation properties of the SP and MSP populations, which affect flux density measurements, are related to the dispersion measures (DMs) of the pulsars in each population. Most known MSPs are at relatively nearer distances compared to SPs, and the DMs of MSPs are on average smaller than for the SPs.
The 16\%, 50\%, 84\% quantiles of the 189 MSP DMs are 15.4\,pc\,cm$^{-3}$, 44.7\,pc\,cm$^{-3}$, 124.6\,pc\,cm$^{-3}$, for the 1170 SPs the respective values are 54.3\,pc\,cm$^{-3}$, 160.6\,pc\,cm$^{-3}$, 385.7\,pc\,cm$^{-3}$.
To minimize the effect of scintillation \citetalias{Spiewak2022} base their tabulated Census flux density values on averages of at least 6 and up to 144 observation epochs.
For the SP census, \citetalias{Posselt2023} obtained values for individual epochs of observations covering at least 1000 pulses, except for the pulse widths which were measured on averaged data of multiple epochs as described by \citet{Posselt2021}. SPs require longer observations for statistically meaningful average pulse profiles, and only a fraction of these pulsars is monitored with multiple MeerKAT observations in the framework of the Thousand-Pulsar-Array (TPA) program (e.g. \citealt{Song2023}).
\citetalias{Posselt2023} considered possibly scintillating sources in the further analysis of the Census SPs by estimating a pseudo modulation index within the MeerKAT frequency band.\\ 

In addition to the one-epoch vs multiple-epoch approach  for flux density measurements (and respectively determined spectral indices), there are the following small differences in the measurement methodology. 
The regions of on-pulse for the flux measurements were determined visually for the MSPs, and via a Gaussian-Process method for the SPs.  
For the MSPs, normalised flux density values at 1400\,MHz were obtained from weighted power-law fits in 8 sub-bands of the total 775\,MHz bandwidth \citepalias{Spiewak2022}, whereas the SP fluxes were provided at a centre frequency of 1264~MHz in \citetalias{Posselt2023}. For the purposes of comparing luminosities in the following, we have rescaled the SP data to 1400~MHz using either the measured power law spectral index, or the average spectral index of the population where a measurement is missing. The linear polarisation for MSPs and SPs is determined in a slightly different way, with \citetalias{Posselt2023} applying a lower limit of 0 following \citet{Everett2001}, while \citetalias{Spiewak2022} chose not to do that.

\section{Comparisons of the catalogue parameters}\label{results}
\subsection{The flux density spectra}
\begin{figure}
\includegraphics[width=\columnwidth]{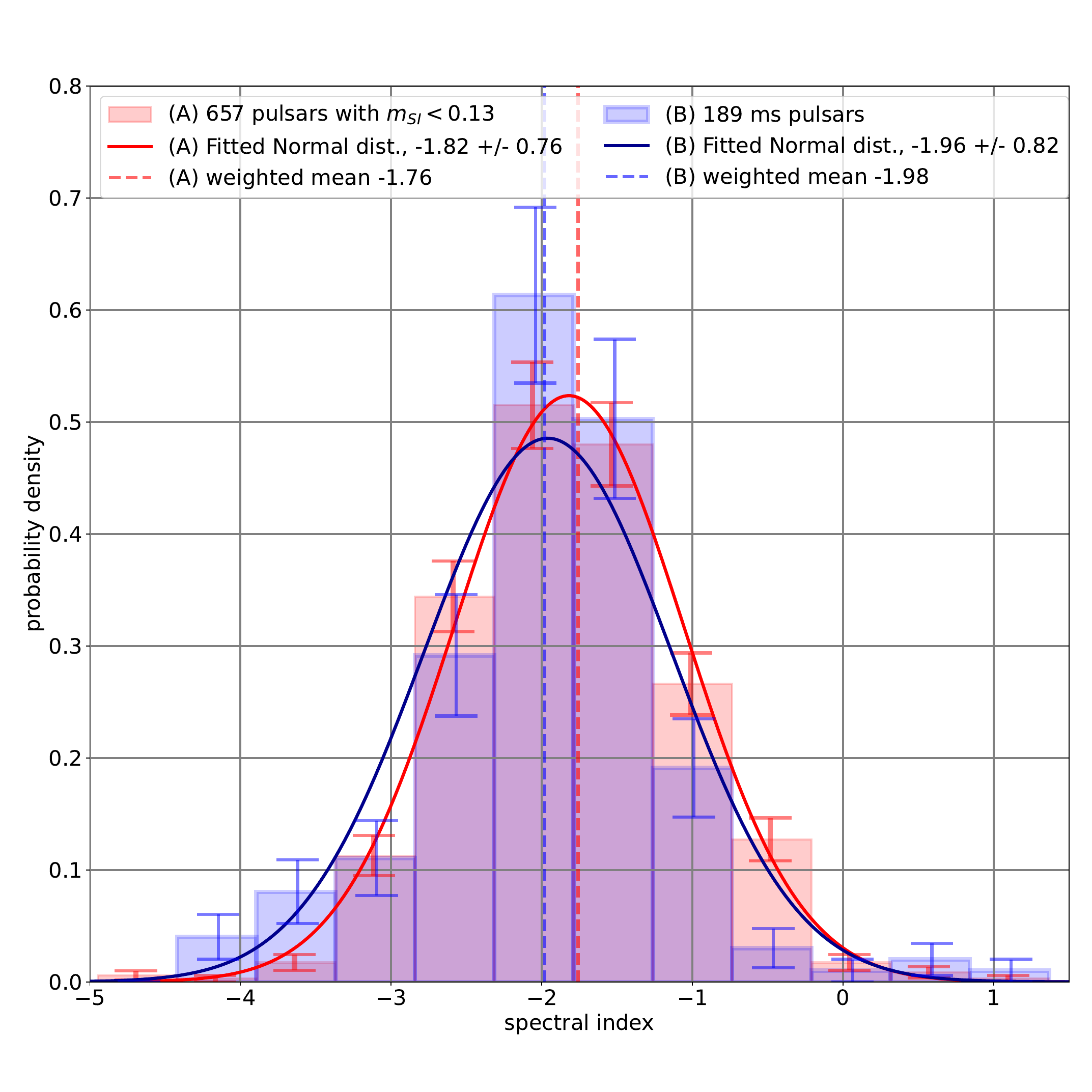}
\vspace{-0.5cm}
\caption{The distribution of the spectral index values for the SPs (red; based on the TPA sample) and the MSPs (blue). The SPs are filtered for low pseudo-modulation index ($m_{\rm SI}$) to minimize the effect of scintillation as in \citetalias{Posselt2023}. Both histograms employ the same value ranges and 20 bins.  The (scaled) Poisson noise of each bin is shown with corresponding colours. The fitted Normal distribution functions for SPs and MSPs are shown with solid lines, the \emph{weighted} means of each distribution are indicated by the respective dashed vertical lines.\label{fig:SIdist}}
\end{figure}

The spectral shape of pulsar's radio flux densities are typically described with a power law $S_{\nu} \propto \nu^{\alpha}$ with spectral index $\alpha$, e.g. \citet{jankowski18}.
We used tabulated values of $\alpha$ for the SPs from \citetalias{Posselt2023}, and those from \citetalias{Spiewak2022} for the MSPs.
Assuming Gaussian distributions for separate fits, we obtain
the location of the maximum at $-1.96$ ($-1.82$) with a standard deviation of 0.82 ($0.76$) for the sample of 189 MSPs (657 SPs). 
\citetalias{Spiewak2022} reported $-1.92$ with a standard deviation of 0.6 for the same MSP data set.
Using 20 equal-width bins in spectral index $\alpha$, Fig. ~\ref{fig:SIdist} shows the SP and MSP $\alpha$-distributions to be similar with maxima and shape agreeing within uncertainties, consistent with the earlier findings of \cite{kramer98}. 

%There is an indication of a slightly skewed MSP distribution that rises slower from the most negative $\alpha$ towards the distribution maximum and then decreases faster (the two bins at $\alpha \sim -4.5$ to $-3.5$, and the bin at $\sim -0.5$ in Fig. ~\ref{fig:SIdist}) compared to the SPs. This minor effect (within the limits of Poisson noise) persists for different numbers of histogram bins.\\ 

The agreement can be probed using the two-sample Kolmogorov-Smirnov (KS) test, for which we estimate a (null hypothesis) probability of $p_{\rm KS}=0.17$ that the two $\alpha$-samples are drawn from the same parent distribution. 
Since the KS-test is insensitive to differences at the wings of the compared distributions\footnote{\href{https://asaip.psu.edu/articles/beware-the-kolmogorov-smirnov-test/}{https://asaip.psu.edu/articles/beware-the-kolmogorov-smirnov-test/}}, we also employ the two-sample non-parametric Anderson-Darling (AD) test
%\footnote{We used the \texttt{scipy.stats.anderson$\_$ksamp} implementation of the AD-test which is capped at a lowest value of 0.1\%.} 
and determine a $p_{\rm AD}=0.048$. This value is almost exactly on the limit ($p_{\rm AD}=0.05$) for rejecting the hypothesis that the distributions arise from the same parent, making the test inconclusive. In summary, the spectral index distributions of MSPs and SPs in the MeerKAT frequency band are broadly similar and our tests do not rule out the assumption that they originate from the same parent distribution. \\

\subsection{Pulse widths}
From the SP and MSP census papers, we use the published pulse widths $W_{10}$ at the 10\% level of the pulse peak.
The $W_{10}$ for the SPs are taken from TPA tables in \citetalias{Posselt2023}, together with the respective spin periods. % important because of the corrected Ps
For the MSPs, we analysed the (one-epoch) pulse profiles of the MSPs by \citetalias{Spiewak2022} that are available on Zenodo. Applying the same methodology as previously done for the TPA, we first determined a smooth noiseless profile with a Gaussian Process before the actual width measurement (see \citealt{Posselt2021} for details). A caveat in using $W_{10}$ measurements from MSP profiles to understand the width of the radio beam arises from the presence of low-level components, that may sit beneath the 10\% threshold. We are exploring techniques to measure low-level MSP components and more precise MSP pulse widths in a future work. \\
%Python GP-package \texttt{George}\footnote{\href{https://george.readthedocs.io}{https://george.readthedocs.io}} by \citet{georgepython}.

\begin{figure}
\includegraphics[width=\columnwidth]{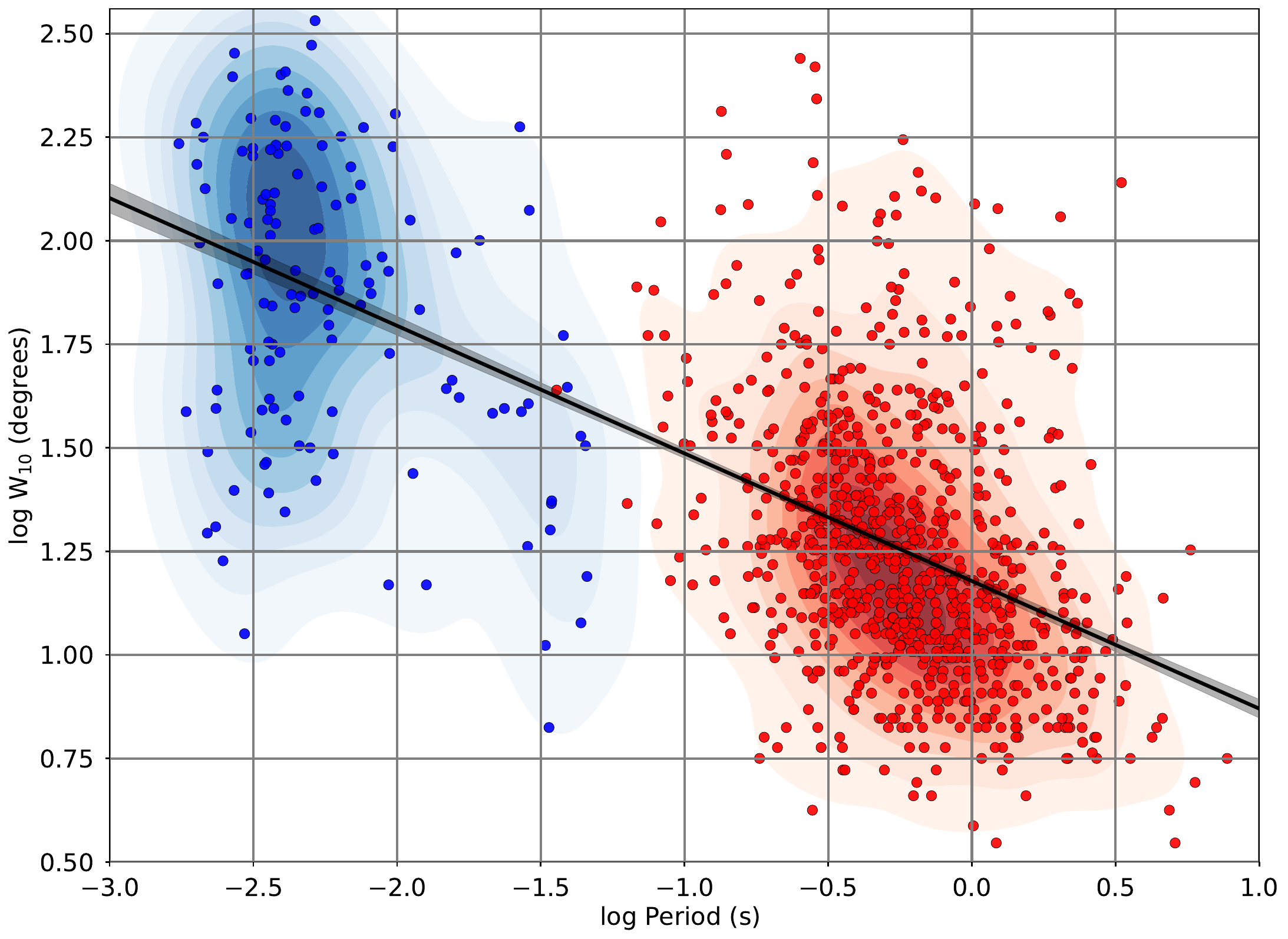}
\vspace{-0.5cm}
\caption{Profile widths $W_{10}$ vs. period for the total bandwidth data in logarithmic space.
The 1010 width measurements consist of 878 values for the SPs from the TPA census (red), and 132 values for the MSPs (blue). Each sub-population is also visualised by a respective density plot. The black line and shaded area show the result of an OLS-fit and its uncertainties, obtained by using bootstrap.
The slope of the OLS-line is $-0.308 \pm 0.014 $. \label{fig:w10pfit}}
\end{figure}

Investigating the period--pulse width relationship for the combined data set of MSPs and SPs (1010 pulsars), we assume that the width is a function of the rotation period $P$, described by a power law with power law index $\mu$. We follow the bootstrapping procedure for a straight-line fit in logarithmic space by \citet{Posselt2021}. In particular, we considered up to 100,000 random samples with replacements to study the distribution of the resulting fit parameters, its general shape, and the 16\%, 50\%, and 84\% quantiles.
Our ordinary least-square fit (OLS) has a Gaussian-shaped distribution of the fit parameters.
We obtained a slope (or PL-index) of $\mu = -0.308 \pm 0.014$ (amplitude of $21.43\degr \pm 0.46$ at $\log P = -0.495$). The fit result is shown in Fig. ~\ref{fig:w10pfit}, together with the data.
Our result for the slope across the \emph{whole} known pulsar population is similar to previous SP results, e.g., the OLS fit results in \citet{Posselt2021} for the TPA pulsars ($\mu =-0.29 \pm 0.03$), and 1.4\,GHz results by \citet{Johnston2019} ($\mu = -0.28 \pm 0.03$). The results do not change by selecting random subsamples of SPs with the same size as the MSPs.

\begin{figure*}%
    \centering
    \subfloat{{\includegraphics[width=8.5cm]{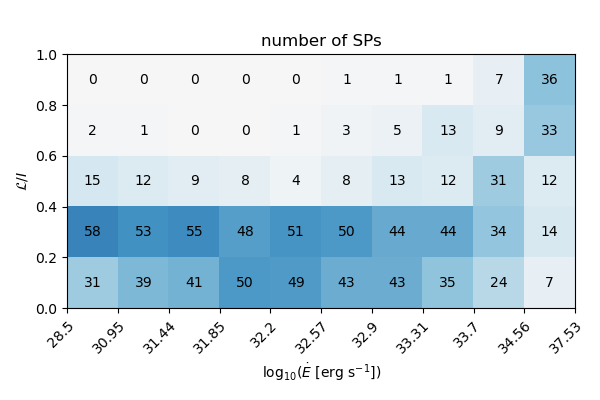} }}%
    \qquad
    \subfloat{{\includegraphics[width=8.5cm]{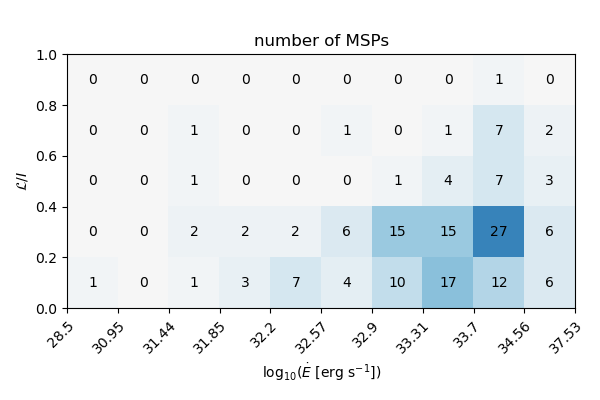} }}%
    \caption{Left: 2D histogram of the number of SPs versus degree of linear polarization $\mathcal{L}/I$ and $\dot{E}$, for bins in $\dot{E}$ that contain approximately equal numbers of pulsars. Right: Same 2D histogram for MSPs, using the same $\dot{E}$ bin definitions as on the left.}%
    \label{fig::lvedot}%
\end{figure*}
\begin{figure}
\includegraphics[width=\columnwidth]{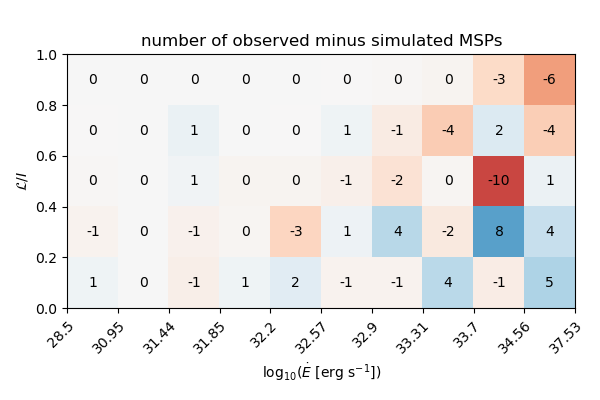}
\vspace{-0.5cm}
\caption{2D histogram of the difference in the number of MSPs observed versus MSPs simulated to match the distribution of SPs in each bin of $\dot{E}$. \label{fig:diffLEdot}}
\end{figure}
\begin{figure}
\includegraphics[width=\columnwidth]{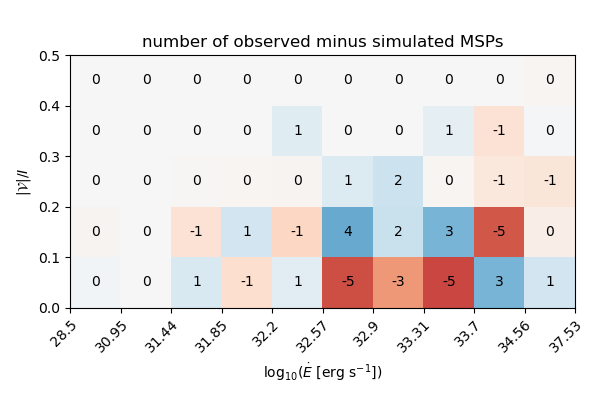}
\vspace{-0.5cm}
\caption{Same as Fig. \ref{fig:diffLEdot} for the fractional absolute circular polarization $|\mathcal{V}|/I$. \label{fig:diffVEdot}}
\end{figure}
\subsection{Linear polarization and $\dot{E}$}
One way of testing the similarities in linear polarization between SPs and MSPs is to compare the relationship of fractional linear polarization $\mathcal{L}/I$ with $\dot{E}$, where $\mathcal{L}$ is the standard de-biased linearly polarized power across the pulse (from averaging Stokes $Q$ and $U$) and $I$ the total power. From the SPs in \citetalias{Posselt2023}, we create a 2D histogram in $\mathcal{L}/I$ and $\dot{E}$ space, such that there are approximately equal numbers of pulsars per $\dot{E}$ bin, summed across all $\mathcal{L}/I$ bins, as shown on the left of Fig. \ref{fig::lvedot}. The $\mathcal{L}/I$ bins are of equal width, 5 bins from 0.0 to 1.0. Using the same bin edges, we generate the same 2D histogram for the MSPs, seen on the right of Fig. \ref{fig::lvedot}. Each $\dot{E}$ column $i$ in the MSP histogram then contains $N_i$ sources. Having converted the corresponding $\mathcal{L}/I$ histogram of the SPs into a probability density function, we draw $N_i$ samples for each $\dot{E}$ column bin. Following this procedure for all MSP $\dot{E}$ bins $i$, we simulate a population of MSPs that follows the $\mathcal{L}/I$ statistics of the SPs. Repeating the simulation 1000 times gives us a mean for each bin of the simulated 2D histogram. Fig. \ref{fig:diffLEdot} shows the difference between the observed MSP 2D histogram, and the simulated MSP 2D histogram generated under the assumption that MSPs follow the same distribution of $\mathcal{L}/I$ as SPs. 

Figs. \ref{fig::lvedot} and \ref{fig:diffLEdot} show a common trend in the two populations of larger numbers of highly polarized pulsars at higher $\dot{E}$. It is worth noting that the simulated $\mathcal{L}/I$ is in general higher than the observed, most notably for MSPs with the highest $\dot{E}$.

We have conducted the same analysis for circular polarization, and specifically for the absolute circular polarization as measured in the SP and MSP MeerKAT census papers. The equivalent of Fig. \ref{fig:diffLEdot} for absolute circular polarization is shown in Fig. \ref{fig:diffVEdot}, in which we see some bins with a larger deviation, but an overall noisy pattern that does not suggest systematic differences.

\subsection{Rotating vector model}
In \citet{paper11} we applied the rotating vector model (RVM) to 1267 SPs from \citetalias{Posselt2023}. Pulsars with scattered profiles and those with low linear polarization were rejected from the sample. The remainder were entered into a fitter and the results of the fitter resulted in a classification of the position angle swing into three classes. We found that 50\% of the pulsars were in the `RVM' class, 8\% of the pulsars had a very shallow slope of PA and were in the `FLAT' class and the remaining 42\% were classified as `non-RVM'.

We applied the same fitting process to 137 MSPs from \citetalias{Spiewak2022}. The output classification resulted in 69 pulsars (50\%) in the `RVM' class, 14 pulsars (8\%) in the `FLAT' class and 54 pulsars (42\%) in the `non-RVM' class. Fig. 5 of \citet{paper11} shows that for SPs, there is a clear difference in the $\dot{E}$ distributions of the three classes. The same conclusion cannot be reached for the MSPs considered here, where we see no obvious difference. The overall $\dot{E}$ distribution of MSPs is much narrower than SPs, however, and this population of MSPs is much smaller than the SPs in \citet{paper11}, so we do not wish to place much weight on this conclusion.  The fraction of MSPs with flat PAs is also somewhat in tension with the relatively high fraction ($\sim33$\%) of flat PAs reported for 24 relativistic binary MSPs in \cite{relbin}. 

In summary, the methodology of \citet{paper11} applied to SPs and MSPs results in a remarkably similar classification of RVM-line, non-RVM, and FLAT PA profiles for the two populations. However, further scrutiny must be applied to the relativistic binary MSPs to check if their reported flat PA profiles present a different picture, and if so, why.

\subsection{Luminosity}
The radio luminosity of pulsars is reported on and referred to in a variety of ways in the relevant literature. It is useful to include here a brief glossary of terms, their definitions, and properties.

{\bf Radio luminosity, $L_R$}. This term we will use for the true time-averaged power in the radio emission mechanism, across all frequencies where emission is present, and the entire solid angle that it covers. This quantity is not accessible observationally, as our line of sight to a pulsar only samples a 1D cut of the radio beam.
\begin{figure*}%
    \centering
    \subfloat{{\includegraphics[width=8.5cm]{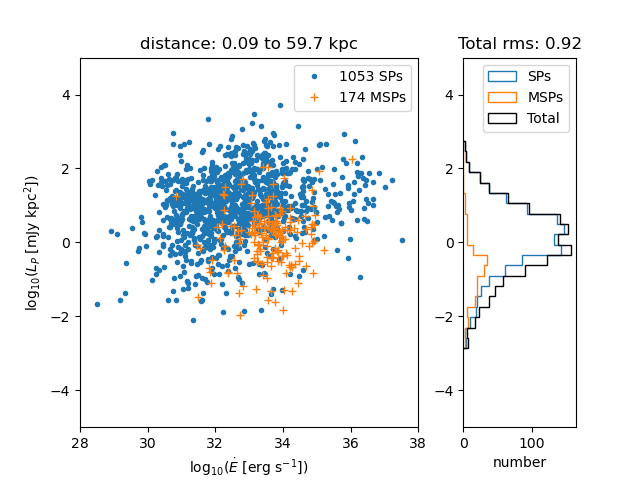} }}%
    \qquad
    \subfloat{{\includegraphics[width=8.5cm]{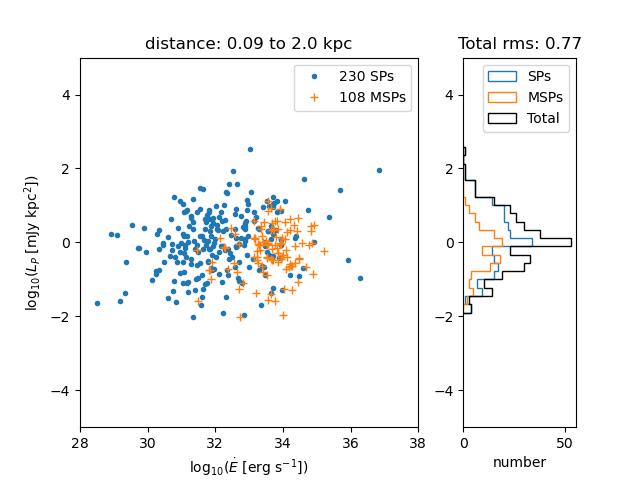} }}%
    \caption{The pseudo luminosity $L_P$ in mJy~kpc$^2$, plotted against the spin down energy $\dot{E}$, accompanied by a histogram of the residual in $L_P$ after subtracting the best fit power-law to the combined MSP and SP $L_P$-$\dot{E}$ data. The standard deviation of the residual of the total population is given numerically above the histogram (rms). SPs are shown in blue dots and MSPs in orange crosses. On the right, only pulsars nearer than 2~kpc are shown, as opposed to the entire list of MeerTIME MSPs and SPs.}%
    \label{fig::L_P}%
\end{figure*}
\begin{figure*}%
    \centering
    \subfloat{{\includegraphics[width=8.5cm]{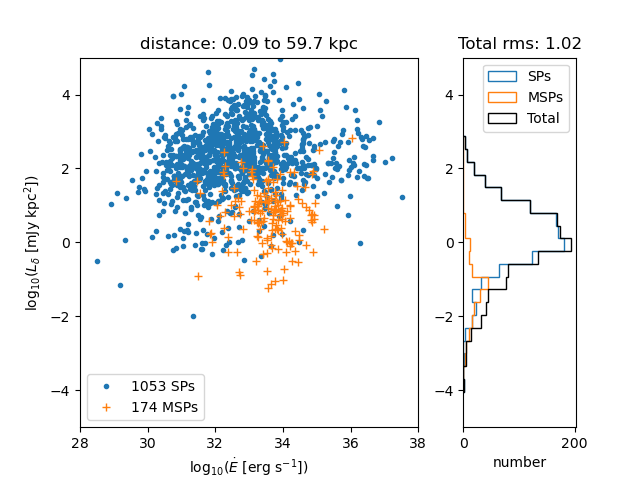} }}%
    \qquad
    \subfloat{{\includegraphics[width=8.5cm]{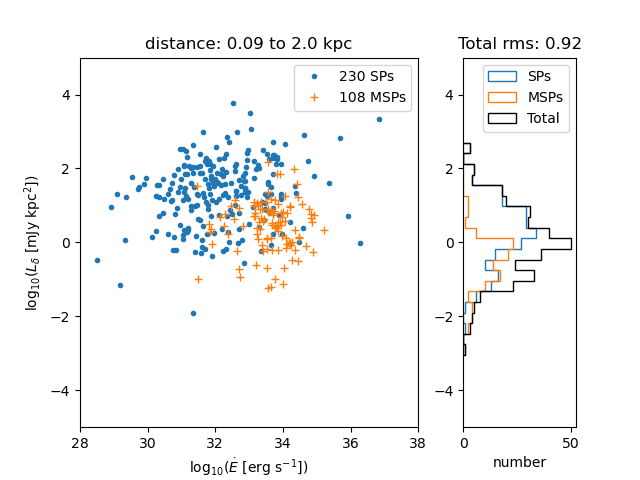} }}%
    \caption{Same as figure \ref{fig::L_P}, but this time showing $L_\delta$, i.e. the pseudo-luminosity corrected by duty cycle. Notice how the populations of MSPs and SPs separate further compared to Fig. \ref{fig::L_P}, even for the nearby population. }%
    \label{fig::L_delta}%
\end{figure*}

{\bf Pseudo luminosity, $L_P$}. This is an observable quantity. It is useful to limit the use of this term to the quantity $L_P = \Smeanm~d^2$, where  \Smean is the mean flux density within a certain observing radio frequency band centred at $\nu$, and $d$ the distance to a pulsar. The pulsar catalogue \textsc{psrcat} \citep{psrcat} provides values of $L_P$ for 400 and 1400~MHz, and distances. For the distance, we use the catalogue value of the \texttt{Dist} parameter, which adopts the YMW16 DM-distance \citep{YMW16} unless there is a measurement believed to be more reliable than the DM-derived distance, or a measured annual parallax. Adopting the YMW16 distance for all sources has almost no effect on the following results. Adopting the older NE2001 DM-distance \citep{NE2001}, does not change any of the following results qualitatively, despite the overall distribution of luminosities being narrower. This is merely a reflection of the difference in the distance distributions between NE2001 and YMW16. Knowing the distance to each pulsar is essential to estimating the luminosity of a given pulsar. However, for a population analysis such as the one presented here, the main consideration is that the uncertainties in the distance to each source are not correlated with other parameters across the population.

In an interferometric image, \Smean simply corresponds to the observed flux density of a pulsar. In the time domain, \Smean is computed by integrating the flux density over all time samples of the average pulse profile, and dividing by the number of time samples across pulse period. For example, a pulse profile $W$ degrees wide, with an integrated flux density under the pulse of $S_{int}=X$~mJy, has $\Smeanm = X~W/360$~mJy. Given the period-width relationship discussed above, \Smean will therefore statistically depend on the pulse period $P$. 

\begin{figure*}%
    \centering
    \subfloat{{\includegraphics[width=8.5cm]{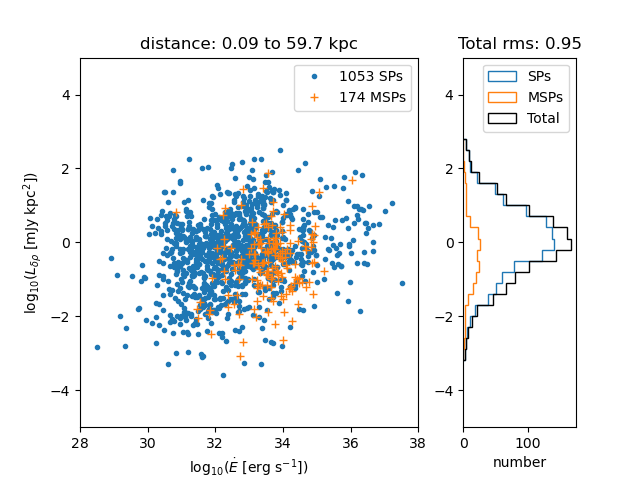} }}%
    \qquad
    \subfloat{{\includegraphics[width=8.5cm]{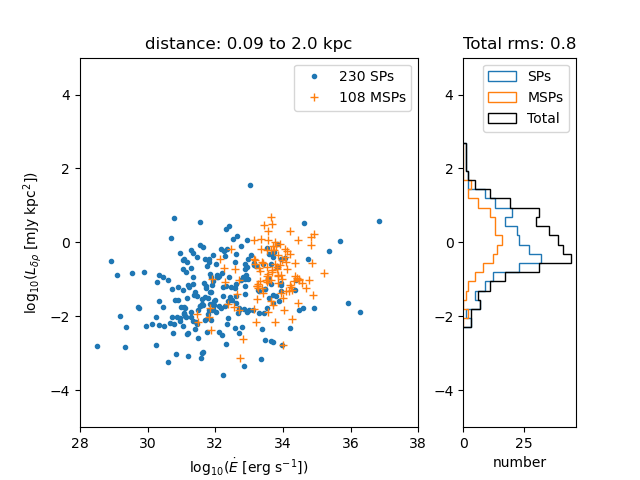} }}%
    \caption{Same as figure \ref{fig::L_P}, but this time showing $L_{\delta\rho}$, i.e. the pseudo-luminosity corrected by duty cycle and beaming angle, using the pulse width as a proxy for the angular size of the beam. This brings the SP and MSP populations closer, albeit with a broader overall distribution than $L_P$.}%
    \label{fig::L_deltarho}%
\end{figure*}

\begin{figure*}%
    \centering
    \subfloat{{\includegraphics[width=8.5cm]{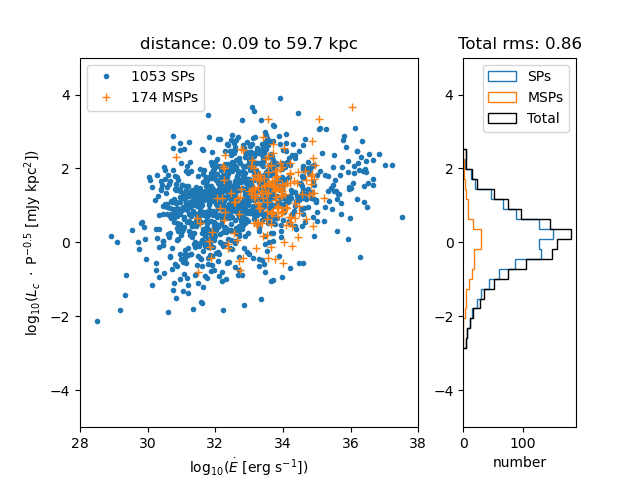} }}%
    \qquad
    \subfloat{{\includegraphics[width=8.5cm]{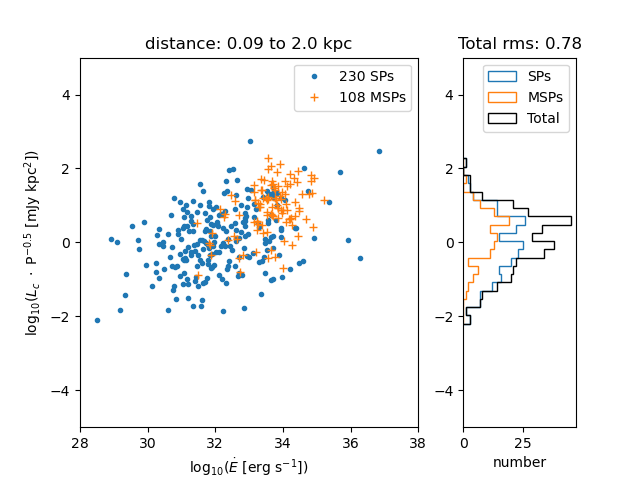} }}%
\\
    \centering
    \subfloat{{\includegraphics[width=8.5cm]{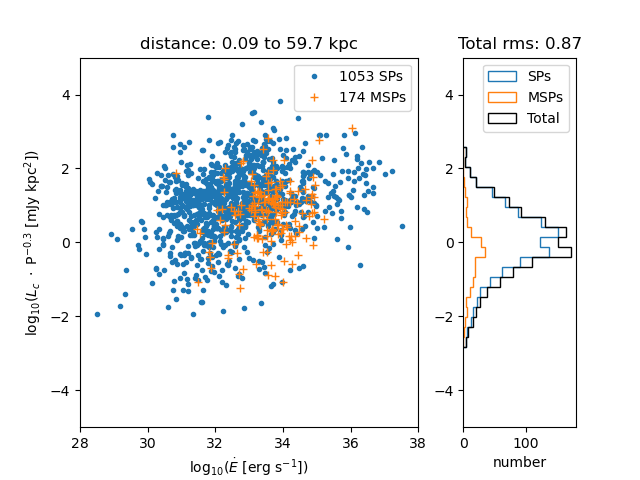} }}%
    \qquad
    \subfloat{{\includegraphics[width=8.5cm]{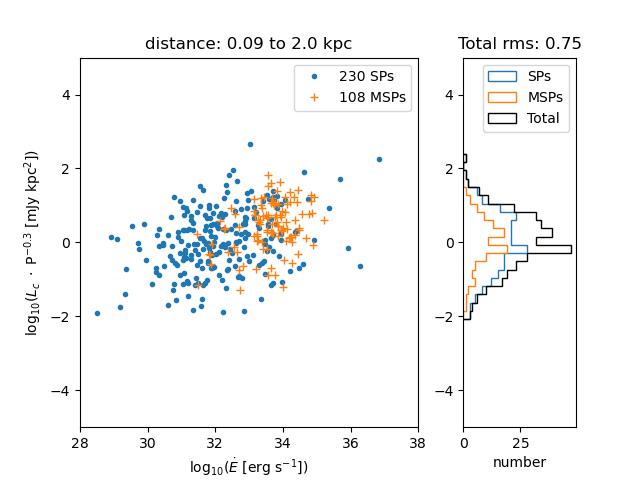} }}%
    \caption{Same as figure \ref{fig::L_P}, but this time showing $L_C$, i.e. the pseudo-luminosity corrected by duty cycle and beaming angle, for the case where the width scales as $P^{-0.5}$ (upper row of plots) and for the case where the width scales as $P^{-0.3}$ (lower row of plots). The histogram of the residuals of the data minus the model given in Eq. 4, using the best fit values of $h$ and $\alpha$}.%
    \label{fig::L_C}%
\end{figure*}

This thinking puts into question the use of \Smean in the context of determining the radio luminosity, especially when drawing conclusions regarding a large population. \Smean scales the pulsed radio emission to provide a measure of the flux density of a pulsar as a continuum source. But as a pulsar spins, the radio beam is often pointing away from the telescope. The example above illustrates how using \Smean to compute $L_P$ is as if the pulsar stops radiating when the radio beam points away. $L_P$ has been used extensively in the literature \citep[e.g.][]{hitrunNG,hitrunCameron}, and here Fig.  \ref{fig::L_P} shows the logarithm of $L_P$ for the MSPs and SPs for which we have measurements, plotted against $\dot{E}$. In \citetalias{Posselt2023} and \citetalias{Szary14} the pseudo luminosity is found to be correlated with the spin-down luminosity $\dot{E}$, specifically with a power-law dependence. The histograms in Fig.  \ref{fig::L_P} and subsequent figures are generated by subtracting a power-law fit with $\dot{E}$ from the data, details of which we present in Section \ref{scaling}. The parameters of the fit are given in Table \ref{tab::fitresults}.
For each of Figs. \ref{fig::L_P} to \ref{fig::L_C}, the left figure shows the entire population, whereas the figure on the right shows only those pulsars with a distance up to 2~kpc. Two features of the left figure are striking, and have been seen in previous works, namely the offset between the SP and MSP populations \cite[e.g.][]{Szary14}, and the shallow slope of $L_P$ versus $\dot{E}$ (e.g. \citetalias{Posselt2023}).  

Motivated by \cite{kramer98}, if we restrict ourselves to the pulsars within a distance of 2~kpc, where scattering and dispersion smearing are less likely to limit the pulsars we can discover, neither of the above two features are as obvious. The right plot of Fig. \ref{fig::L_P} shows that the nearby population of 230 SPs (compared to a total of 1053) and 108 MSPs (compared to a total of 174) have $L_P$ distributions that appear more similar to each other than the histograms of the entire population, shown on the left of Fig. \ref{fig::L_P}. There is no doubt that understanding luminosity critically depends on understanding distances and the systematic effects introduced by the sensitivity of surveys. We discuss this in Section \ref{distance}. First we consider how the duty cycle and beaming fraction complicate the picture further. 

{\bf Scaled pseudo luminosity, $L_S$}. The problem with using $L_P$ defined as above, is addressed in the literature by scaling the pseudo luminosity to account both for beaming, and for the fact that the line-of-sight cut through the beam is one dimensional. This scaled pseudo luminosity is defined as $L_S=A~L_P$. Leaving the flux density spectrum to the side for now, $A$ has two terms: $A = 1/\delta~4\pi\sin^2\rho/2$ \cite[see][]{handbook}, where $\delta$ is effectively the duty cycle $W/360$ term from above, converting \Smean into $S_{int}$. The second term multiplies by the solid angle of the beam to account for the beam opening angle $\rho$, defined as the angle between the tangent of the last emitting field lines and the magnetic axis, at the emission height. This latter term assumes a uniformly illuminated beam. The assumption is difficult to justify entirely, but also difficult to replace with some other. Given the statistical similarities in the spectra between SPs and MSPs shown in Fig.  \ref{fig:SIdist}, we do not consider the spectral index further as a useful parameter for this discussion, as there are more important corrections to consider first.

$L_S$ is usually computed as an approximation to $L_R$, for comparisons with the spin-down luminosity $\dot{E}$. It is, however, often computed for a population of pulsars, using a single scaling constant of $7.4\times10^{27}$ \cite[][e.g.]{Szary14, Posselt2023} for the entire population, despite the assumptions for the constant being $\delta\approx0.04$, $\rho\approx 6\degree$, a fixed spectral index of $-1.8$ and an observing frequency of 1400~MHz. Using $L_{S*}$ as $7.4\times10^{27}~L_P$, or using the pseudo luminosity $L_P$ itself \citep[e.g.][]{levin13}, the consequences in drawing conclusions about pulsar populations are too important to take lightly. Effectively, the correction from \Smean to $S_{int}$ is not made, as  $\delta$ is kept a constant across the population. This not only introduces a systematic error, it also introduces an error that is statistically dependent on the period $P$, in the way mentioned above. For the second term of $A$ also, this implies no change in the solid angle of the beam across the population. This is again in contradiction with the period width relationship, assuming the pulse width is in some way representative of the beam opening angle. 

\section{Scaling the radio pseudoluminosity}\label{scaling}
Considering the two quantities, $L_P$ and $L_{S*}$ defined with constant $A$, we can define a third, $L_\delta$, namely $L_P/\delta$, which under the assumption of a fully illuminated beam, corrects for the measurable duty cycle, and plot it as before in Fig.  \ref{fig::L_delta}. If we limit ourselves to the nearby population initially (the plots on the right), we see that the $\delta$ correction separates the SPs and MSPs even for nearby sources. One conclusion to draw from this is that it is unlikely that the right hand plot of Fig.  \ref{fig::L_P} can be used to claim that the luminosity function of SPs and MSPs are similar, without further considerations. 
\begin{figure*}%
    \centering
    \subfloat{{\includegraphics[width=8.5cm]{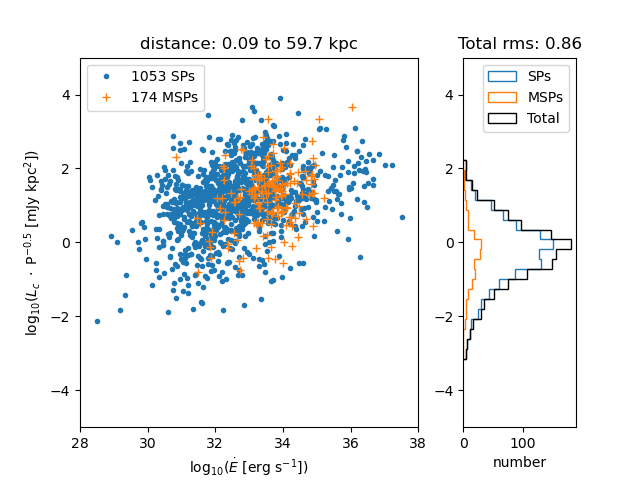} }}%
    \qquad
    \subfloat{{\includegraphics[width=8.5cm]{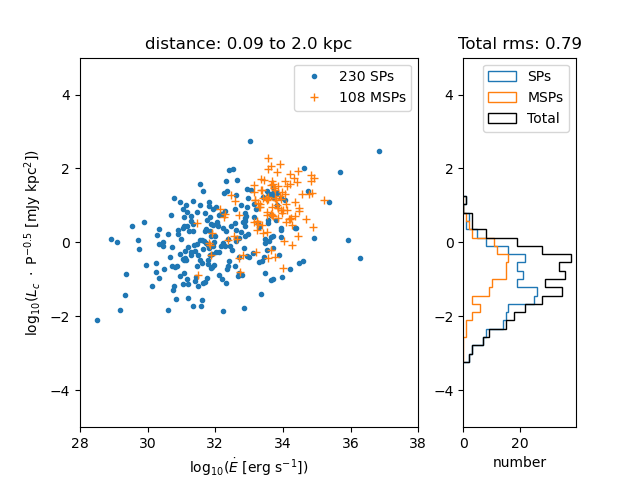} }}%
    \caption{Same as figure \ref{fig::L_P}, but this time showing $L_C$, with $h$ fixed to $-6.1$, $\beta$ fixed to -0.5, and $\alpha$ fixed to 0.23 in both the nearby and the entire population. }%
    \label{fig::L_fixed}%
\end{figure*}

Scaling $L_P$ by $1/\delta$ is informative, but does not account for the angular size of the radio beam. Once again, under the assumption that the pulse width is representative of a uniformly illuminated beam across some group of open magnetic field lines, it is informative to further use $\delta$ to scale for the solid angle the beam is radiating into. The simplest way to do this, with all associated caveats, is to assume the beam solid angle scales with $\delta^2$, in other words $\delta$ is a measure of the diameter of a 1D cut of the beam and $\delta^2$ is a measure of the area of the 2D beam. Of course geometry is critical in each individual case, and geometry may change in a systematic way across the population. Nevertheless, it is informative to proceed. For this purpose we define $L_{\delta\rho}\equiv~L_P~(1/\delta)~(\delta^2) = \delta~L_P$. Fig.  \ref{fig::L_deltarho} shows the results of this scaling. The effect is that the SP and MSP populations draw nearer to each other, but the overall spread increases. The latter is not surprising given the assumptions.  

What if the power-law dependence on $\dot{E}$ holds true for $L_R$? We have shown that $L_{S*}$, $L_\delta$, and $L_{\delta\rho}$ have limitations as proxies for $L_R$. Here, we adopt $L_P$ as our starting point, like \citet{kramer98}. One approach to scaling $L_P$ in order to comment on $L_R$ comes from making the hypothesis that both terms of $A$ are functions of the pulse period, and therefore $A$ itself is some function of the pulse period. We define $A_P \equiv K~P^{\beta}$, and $L_C \equiv A_P~L_P$. If $L_C$ is a good proxy of $L_R$, then our assumption above means that:
\begin{equation}
L_C \equiv A_P~L_P = C~\dot{E}^{\alpha},      
\end{equation}
so
\begin{equation}
K~P^{\beta}~L_P = C~\dot{E}^{\alpha}     
\end{equation}
and, combining the constants $K$ and $C$ together,
\begin{equation}
L_P = H~\dot{E}^{\alpha}~P^{-\beta}.    
\end{equation}
Taking the logarithm of this gives us:
\begin{equation}
\log(L_P) = h + \alpha~\log(\dot{E})-\beta~\log(P)    
\label{eq::scaling}
\end{equation}

This is an equation with two independent variables, namely $\dot{E}$ and $P$, and three parameters, namely $h$, $\alpha$, and $\beta$.

At this point, we have 2 choices for the value of $\beta$. The first is to consider whether our definition of $A$ and the dependencies of its two terms on $P$ can inform us. As mentioned above, $\delta$ will scale as the period-width relationship, which has published values in the literature of $-0.5$ and approximately $-0.3$ for the exponent of $P$ \citep[e.g.][]{Johnston2019}, therefore $1/\delta$ may scale as $P$ to the $0.5$ or $0.3$. The polar cap size, as a proxy for the beam solid angle, may scale as $P$ to the $-1.0$ or $-0.6$. Combining these two terms together results in values of $\beta$ of either $-0.5$ or $-0.3$, reiterating the assumption that the beam is both circular and entirely filled, which is a topic of active investigation.   The two rows of Fig.  \ref{fig::L_C} show $L_C$ for the two aforementioned scalings of $\beta$ with the $P$. In the plots that cover all distances, the MSP and SP populations are slightly separated as in Fig. \ref{fig::L_deltarho}, but the residual histograms after removing the best model in the form of Eq. \ref{eq::scaling} are tighter than Fig. \ref{fig::L_deltarho}, as shown by the measured rms. The same holds true for the nearby populations up to 2~kpc in the plots on the right, see Table \ref{tab::fitresults}.

\begin{table}
\centering
\caption{Maximum likelihood values for fits of Eq. \ref{eq::scaling}, given here for completeness and to help with comparisons.}
%\begin{tabular}{l{3cm}L{1.4cm}L{1.4cm}L{1.4cm}}
\begin{tabular}{lLLLl}
\hline
\textbf{scaling of $L_P$} & \textbf{$\alpha$} & \textbf{$h$} & \textbf{$\beta$} & \textbf{rms} \\
\hline
Fig. \ref{fig::L_P}, all distances & 0.09 & -2.01 & 0.0 & 0.92\\
Fig. \ref{fig::L_P}, distance$<2$~kpc & 0.07 & -2.4 & 0.0 & 0.77\\
Fig. \ref{fig::L_delta}, all distances & 0.02 & 1.48 & 0.0 & 1.02\\
Fig. \ref{fig::L_delta}, distance$<2$~kpc & -0.06 & 3.12 & 0.0 & 0.92\\
Fig. \ref{fig::L_deltarho}, all distances & 0.16 & -5.49 & 0.0 & 0.95\\
Fig. \ref{fig::L_deltarho}, distance$<2$~kpc & 0.21 & -7.92 & 0.0 & 0.80\\
Fig. \ref{fig::L_C} top, all distances & 0.23 & -6.45 & -0.5 & 0.86\\
Fig. \ref{fig::L_C} top, distance$<2$~kpc & 0.33 & -10.29 & -0.5 & 0.78\\
Fig. \ref{fig::L_C} bottom, all distances & 0.18 & -4.68 & -0.3 & 0.87\\
Fig. \ref{fig::L_C} bottom, distance$<2$~kpc & 0.22 & -7.14 & -0.3 & 0.75\\
Fig. \ref{fig::L_fixed} all distances & 0.23 & -6.1 & -0.5 & 0.86\\
Fig. \ref{fig::L_fixed} distance$<2$~kpc & 0.23 & -6.1 & -0.5 & 0.79\\
\hline
\label{tab::fitresults}
\end{tabular}
\end{table}

%Results of the fit and figures
\section{Radio luminosity and distance}\label{distance}

Figs. \ref{fig::L_P} to \ref{fig::L_C} suggest that no matter what scaling we apply to $L_P$, the MSP and SP populations separate when considering sources at all distances, while they appear more or less well matched for nearby sources up to 2~kpc. Assuming some intrinsic distribution of the radio luminosity that is common between MSPs and SPs, and given some limits to the sensitivity of pulsar surveys, the nearby population should be sampling the luminosity distribution more completely than the distant population, as long as there are sufficient numbers of nearby sources. That is to say, we can better detect intrinsically faint pulsars when they are nearby, and we are likely to be sampling more of the intrinsically brighter population from further away. The volume of Galaxy also increases with distance, meaning the luminous and distant pulsars are favoured twice: there are more of them, and their less luminous counterparts are too weak to find. 

Even in $L_C$, the MSP and SP populations separate to some degree at larger distances, so one question that arises is where are the very luminous and distant MSPs? There are at least two possible answers. The first is that the population of nearby known MSPs corresponds to those sources with the highest intrinsic radio luminosity $L_R$. This would go against the previously published claims that we are missing high-luminosity MSPs from the nearby population, as stated in \cite{kramer98}. That would also mean that we are unlikely to find significant numbers of distant MSPs, as they would need to be more luminous than the high-luminosity nearby population. The second, is that we are insensitive to distant MSPs for reasons other than luminosity, such as interstellar scattering and dispersion, reasons that are well known limitations in pulsar surveys, discussed in e.g. \cite{levin13}. In this case, high-frequency surveys may indeed reveal larger populations of highly luminous MSPs. \cite{snake} report an interesting example of exactly such a source.

So how can we compare the luminosities of MSPs and SPs, considering the above arguments? There is no doubt the populations of SPs and MSPs studied here are limited by systematics. Among other reasons, this arises because whether a pulsar is found in a survey or not depends on the flux density and pulse width, rather than the intrinsic radio luminosity $L_R$. Pulsars with narrow, bright profiles are easier to discover than pulsars with wide, low signal-to-noise profiles. For the MSPs there are additional obvious and known difficulties in finding them at large distances owing to interstellar propagation \cite[as per e.g.][]{levin13}, which may be limiting the number of high-luminosity sources, or at least preventing us from exploring that region of parameter space. For the SPs, the population is dominated by the bright and distant SPs as argued above. 

Figs. \ref{fig::L_P} to \ref{fig::L_C} reveal one interesting detail, which is especially apparent in Fig. \ref{fig::L_C}. The width of the distribution of the fit residuals for the nearby population is not much narrower than the residual for the entire population, as stated in Table \ref{tab::fitresults}. One may point out that these residuals are obtained for different values of the constants $h$, $\alpha$, and $\beta$ of Eq. \ref{eq::scaling}. As a test, we have generated Fig. \ref{fig::L_fixed} by computing $L_C$ using the same parameters of Eq. \ref{eq::scaling} for both the nearby and entire population, as given in bottom rows of Table \ref{tab::fitresults}. We cannot consider the nearby population of SPs to be representative of the luminosity distribution, as the residual histograms clearly show. Nevertheless, when considering all distances, this figure provides additional evidence that, with appropriate scaling, the luminosities of SPs and MSPs may indeed come from the same parent distribution, contrary to previous publications suggesting that MSPs could be an order of magnitude less luminous than SPs. 
%Most catalogues, including the two we are using here for SPs and MSPs, report \Smean rather than $S_{int}$. The duty cycle $\delta$ is typically not available, however a measurement of the pulse width at the 10\% level of the peak can be used to approximate $delta$ as $W_{10}/P$.  

\section{Conclusions}\label{conclusions}
We have used the current most complete single-survey catalogues for the parameters of non-recycled and recycled pulsars, to draw a detailed comparison between these populations, with an emphasis on their radio emission processes. We have focused on the spectral properties, the pulse width versus rotational period, the polarization, and the luminosity, motivated primarily by \citet{kramer98} and \citet{xilouris98}. We have not investigated potential differences between isolated and binary pulsars, which has been done in the past for unscaled values of $L_P$, as in \cite{burgaybinaries}. In summary:
\begin{itemize}
    \item We have compared the spectral index distribution of SPs and MSPs and find them to agree within the errors.
    \item Using $W_{10}$ widths, we find the SP and MSP population to obey the same period-width relationship, with a slope of approximately $-0.3$ in log-log space. We are investigating the caveat that $W_{10}$ may be insensitive to low-level components in MSPs.
    \item We have compared the fraction of SPs and MSPs that have linear polarization position angle profiles compatible with the RVM, and we have found the fractions to be near identical.
    \item The fractions of SPs and MSPs with relatively flat PAs also seems near identical, with further investigation needed for the MSPs in relativistic binaries. 
    \item We have compared the degree of linear polarization vs $\dot{E}$ in SPs and MSPs, and found that, at low significance, the MSP population shows slightly fewer very highly polarized pulsars at high $\dot{E}$ compared to SPs. Similarly there are no striking differences in the degree of absolute circular polarization versue $\dot{E}$ between SPs and MSPs. 
    \item We have analysed the methodology by which the radio luminosity has been considered in the past and we demonstrate that there are clear shortcomings to using a constant for scaling the pseudo-luminosity across the entire population. It is instead essential to capture the effects of duty cycle and beaming.
    \item We show that MSPs likely follow the same power law of radio luminosity versus $\dot{E}$ as SPs, once the pulse duty cycle and beam solid angle is taken into account. This contradicts past findings that MSPs are less luminous, and has an impact on prioritizing high radio frequencies over sensitivity for survey planning. 
\end{itemize}
So is there a differece in the radio luminosities of SPs and MSPs? To dive deeper into this question, we are preparing a future paper analyzing the average profiles in both populations. Specifically, we are investigating the effect of low-intensity MSP components, which affect our pulse width measurements and hence the scaling of radio luminosity discussed here. We are also investigating component multiplicity and associations based on the average pulse shapes between the two populations. \cite{kj07} showed an increase in the fraction of multi-component profiles with decreasing $\dot{E}$, and a similar analysis for the MSP dataset will be presented in a future paper.

A further next step in uncovering the underlying luminosity distribution is to sample globular cluster MSPs to remove the distance uncertainty. Looking to the future, there are two distinctly interesting possibilities that will affect future analysis. If higher frequency surveys find a population of luminous and distant MSPs, that suggests that the nearby population is not limited by luminosity. On the other hand, if more sensitive telescopes such as the Square Kilometre Array discover a population of lower luminosity, nearby MSPs, that may suggest that indeed the current MSP population is already sampling the high end of the intrinsic luminosity distribution. One way or another, investigating the radio luminosity of a population of pulsars requires scaling the historically used pseudo-luminosity to account for the large variation of the beam size across the population. For now, despite the large differences in inferred \bsurf and \tauc between MSPs and SPs, it cannot be ruled out that the radio emission mechanism has the same efficiency between the two populations. 
\section{Acknowledgements}
The MeerKAT telescope is operated by the South African Radio Astronomy Observatory (SARAO), which is a facility of the National Research Foundation, an agency of the Department of Science and Innovation. SARAO acknowledges the ongoing advice and calibration of GPS systems by the National Metrology Institute of South Africa (NMISA) and the time space reference systems department department of the Paris Observatory. We thank Marcus Lower for comments on the manuscript, and we acknowledge the use of the following Python packages: 
\texttt{ASTROPY}  \citep{astropy}, 
\texttt{MATPLOTLIB} \citep{matplotlib}, 
\texttt{PANDAS} \citep{pandas,reback2020pandas},
\texttt{SCIPY} \citep{2020SciPy-NMeth}, and 
\texttt{SEABORN} \citep{seaborn}.
\section{Data availability}

The data used in this manuscript are published in \cite{Posselt2023} and \cite{Spiewak2022}. 

\bibliographystyle{mnras}
\bibliography{TPAaMSP.bib}

%\begin{thebibliography}{99}
%\end{thebibliography}

%%%%%%%%%%%%%%%%%%%%%%%%%%%%%%%%%%%%%%%%%%%%%%%%%%

%%%%%%%%%%%%%%%%% APPENDICES %%%%%%%%%%%%%%%%%%%%%

%%%%%%%%%%%%%%%%%%%%%%%%%%%%%%%%%%%%%%%%%%%%%%%%%%

% Don't change these lines
\bsp	% typesetting comment
\label{lastpage}
\end{document}